\documentclass[english,aps,manuscript,aps,manuscript,showpacs]{revtex4}
\usepackage[T1]{fontenc}
\usepackage[latin9]{inputenc}
\usepackage{float}
\usepackage{amsbsy}
\usepackage{amstext}
\usepackage{amssymb}
\usepackage{graphicx}

\makeatletter

\newcommand{\noun}[1]{\textsc{#1}}

\@ifundefined{textcolor}{}
{%
 \definecolor{BLACK}{gray}{0}
 \definecolor{WHITE}{gray}{1}
 \definecolor{RED}{rgb}{1,0,0}
 \definecolor{GREEN}{rgb}{0,1,0}
 \definecolor{BLUE}{rgb}{0,0,1}
 \definecolor{CYAN}{cmyk}{1,0,0,0}
 \definecolor{MAGENTA}{cmyk}{0,1,0,0}
 \definecolor{YELLOW}{cmyk}{0,0,1,0}
}

\makeatother

\usepackage{babel}
\begin{document}

\title{Periodic driving control of Raman-induced spin-orbit coupling in
Bose-Einstein condensates: the heating mechanisms}

\author{J.M. Gomez Llorente and J. Plata}

\address{Departamento de F\'{\i}sica, Universidad de La Laguna,\\
 La Laguna E38204, Tenerife, Spain.}
\begin{abstract}
We focus on a technique recently implemented for controlling the magnitude
of synthetic spin-orbit coupling (SOC) in ultra-cold atoms in the
Raman-coupling scenario. This technique uses a periodic modulation
of the Raman-coupling amplitude to tune the SOC. Specifically, it
has been shown that the effect of a high-frequency sinusoidal modulation
of the Raman-laser intensity can be incorporated into the undriven
Hamiltonian via effective parameters, whose adiabatic variation can
then be used to steer the SOC. Here, we characterize the heating mechanisms
that can be relevant to this method. We identify the main mechanism
responsible for the heating observed in the experiments as basically
rooted in driving-induced transfer of population to excited states.
Characteristics of that process determined by the harmonic trapping,
the decay of the excited states, and the technique used for preparing
the system are discussed. Additional heating, rooted in departures
from adiabaticity in the variation of the effective parameters, is
also described. Our analytical study provides some clues that may
be useful in the design of strategies for curbing the effects of heating
on the efficiency of the control methods. 
\end{abstract}

\pacs{67.85.-d, 32.10.Fn, 33.60.+q, 37.10.Gh}

\maketitle

\section{Introduction}

The realization of spin-orbit coupling (SOC) in Bose-Einstein condensates
has opened the way to intense experimental and theoretical work which
has brought about significant advances in the research on ultra-cold
atoms \cite{key-SpielmanNature1,key-SpielmanRepProg,key-DalibardRMP}.
Variations of the original setup, including schemes applicable to
fermionic systems \cite{key-fermionSOC1,key-fermionSOC2}, have been
developed, and, a series of fundamental effects have been realized.
Envisaged technical implications of this extended scenario, in particular,
potential applicability to create novel states of matter, like topological
insulators or nontrivial superfluids, are being explored \cite{key-SpielmanNature1,key-SpielmanRepProg}.
Apart from enriching the dynamics in diverse forms, SOC can be considered
as an element to manipulate it. Especially relevant to the strategies
of control, is the recent realization of a technique to steer in real
time the characteristics of SOC. It was proposed that, in the Raman-laser
coupling setup \cite{key-SpielmanNature1}, the application of a high-frequency
sinusoidal modulation of the laser intensity can be used to convert
the system into an effective undriven counterpart where the adiabatic
variation of the modulation amplitude can serve to control the strength
of the SOC \cite{key-ZhangSciRep}.  The experimental realization
confirmed the applicability of this scheme: the detected response
of the system agreed with the theoretical predictions on the changes
in the SOC strength. The system was also observed to present heating
dependent on the driving frequency \cite{key-SpielmanControl}. Here,
we extend the previously applied theoretical framework to account
for the heating mechanisms. Apart from the practical interest of improving
the approach that supports the control method, our study has fundamental
implications: it deals with the development of models for a complex
scenario with elements of general relevance. Different stages, associated
with  increasing precision in the modeling, can be singled out in
our description. We will proceed by isolating first the essential
components of the heating processes, and, dealing, subsequently, with
specific aspects of the considered setup. A primary difficulty is
rooted in the required treatment of the atomic interactions. Indeed,
as the experimental preparation corresponded to a condensate, departures
from the single-particle dynamics must be evaluated. Further complexity
is introduced by the synthetic coupling of internal and external degrees
of freedom. Although there have been significant advances in the analysis
of the effect of the SOC on the dynamics of a condensate \cite{key-StringariSOCreview,key-StringariSOC1,key-StringariSOC2},
the studies have frequently concentrated on a uniform scenario and
on features associated with the lowest band and with low-energy excitations.
Here, the need of dealing with the inter-band transitions induced
by the (high-frequency) driving implies focusing on additional elements.
The evaluation of the rate of population transfer to excited states
requires explicitly taking into account the harmonic trapping. Moreover,
a detailed description of the dispersion relations of the elementary
excitations in different energy ranges is needed. We must also deal
with damping of the excited states, and, therefore, with the role
 of nonlinear inter-mode coupling. The restriction to a particular
set of SOC parameters will allow us to cope with the system complexity,
and, still, identify basic mechanisms responsible for the observed
features. From the obtained insight, some useful clues to setting
up methods for curbing the effects of heating may be extracted. Beyond
the control objectives, our description opens the way to the design
of strategies for exploring different aspects of the SOC systems. 

The outline of the paper is as follows. In Sec. II, central elements
of our methodology are introduced: through an appropriate unitary
transformation, we set up a perturbative scheme to deal with the sinusoidal
driving of the Raman coupling amplitude. The perturbative parameter
will be shown to be given by the quotient between the modulation amplitude
of the Raman-coupling strength and the modulation frequency. In Sec.
III, a single-particle approach is presented to simulate general characteristics
of the experimental findings. The role of perturbative corrections
to the scenario of control proposed in Ref. {[}7{]} is assessed. In
particular, we focus on the first-order term, which accounts for driving-induced
transfer of population to excited states. Atomic-interaction effects
are tackled in Sec. IV: we deal with ground-state properties, elementary
excitations, and damping. Special attention is given to the dependence
of the heating process on decay of the excitations resulting from
inter-mode coupling. In Sec. V, features specific to the ramp implemented
to prepare the system are analyzed. Finally, some general conclusions
are summarized in Sec. VI.

\section{Control of the spin-orbit coupling}

We consider the realization of SOC in Bose-Einstein condensates reported
in Ref. \cite{key-SpielmanNature1}. There, an appropriate arrangement
of Raman lasers was used to create synthetic SOC, specifically, to
couple one of the components of the external linear momentum of the
atoms $\mathbf{p}$ to an effective ``spin'', i.e., to a two-level
internal system formed by hyperfine states. The setup was composed
by two orthogonally polarized Raman lasers with different frequencies
and propagation directions. The arrangement characteristics were designed
to couple two Zeeman-split internal states, and, in particular, to
make the coupling dependent on the atom external momentum. In this
form, an effective SOC was induced. In a single-particle approach,
and, ignoring the effect of the harmonic confinement, that scenario
is described by the Hamiltonian 

\begin{equation}
H=\left(\frac{\hbar^{2}k_{x}^{2}}{2m}+E_{L}\right)I+\frac{\hbar\Omega_{0}}{2}\sigma_{x}+\left(\frac{\hbar\delta_{0}}{2}+\alpha_{0}k_{x}\right)\sigma_{z}
\end{equation}
where $m$ is the atomic mass, $I$ is the unit matrix, $\sigma_{i}$
($i=x,\, y\,,z$) are the Pauli matrices corresponding to the \emph{pseudospin},
i.e., to the considered effective two-level system, and $k_{x}$ denotes
the quasi-momentum in the coupling direction ($\mathbf{p}=\hbar\mathbf{k}$).
Additionally, concerning the laser characteristics, $\Omega_{0}$
denotes the Raman coupling amplitude, $\delta_{0}$ represents an
adjustable detuning, and $\alpha_{0}$ is the strength of the realized
SOC ($\alpha_{0}=2E_{L}/k_{L}$ where $E_{L}=\hbar^{2}k_{L}^{2}/2m$
is the recoil energy and $k_{L}=2\pi\sin(\theta/2)/\lambda$ is the
recoil momentum; $m$ is the atomic mass, $\lambda$ is the laser
wavelength, and, $\theta$ is the angle between the Raman lasers,
which, here, as in the experiments, is taken to be $\theta=\pi/2$
\cite{key-SpielmanNature1,key-SpielmanPRA}). 

Let us see how the SOC strength can be controlled by modulating the
Raman coupling amplitude in the form 

\begin{equation}
\Omega_{0}\rightarrow\Omega(t)=\Omega_{0}+\Omega_{R}\cos(\omega t).
\end{equation}
Appropriate to analyze the dynamics resulting from the driving is
the application of the unitary transformation

\begin{equation}
U(t)=\exp\left[-\frac{i}{2}\frac{\Omega_{R}}{\omega}\sin(\omega t)\sigma_{x}\right].\label{eq:3}
\end{equation}
The transformed Hamiltonian, given by 

\begin{equation}
H^{\prime}=U^{\dagger}HU-i\hbar U^{\dagger}\dot{U},
\end{equation}
is written, after straightforward algebra \cite{key-Louisell}, as

\begin{eqnarray}
H^{\prime} & = & \left(\frac{\hbar^{2}k_{x}^{2}}{2m}+E_{L}\right)I+\frac{\hbar\Omega_{0}}{2}\sigma_{x}+\nonumber \\
 &  & \left(\frac{\hbar\delta_{0}}{2}+\alpha_{0}k_{x}\right)\left\{ \cos[\zeta(t)]\sigma_{z}+\sin[\zeta(t)]\sigma_{y}\right\} ,
\end{eqnarray}
where 

\begin{equation}
\zeta(t)\equiv\frac{\Omega_{R}}{\omega}\sin(\omega t).
\end{equation}
Now, taking into account the expansion of $\cos[\zeta(t)]$ and $\sin[\zeta(t)]$
in terms of the ordinary Bessel functions $J_{n}(u)$ \cite{key-Grad},
we can rewrite the Hamiltonian $H^{\prime}$ as the sum of different-order
contributions. This approach provides us with a perturbative scheme
where different-order corrections to the scenario of control proposed
in Ref. {[}7{]} can be evaluated. Note that the \emph{small parameter}
in the perturbative approach is determined by $\Omega_{R}/\omega$.
The zero-order term, given by 

\begin{equation}
H_{0}=\left(\frac{\hbar^{2}k_{x}^{2}}{2m}+E_{L}\right)I+\frac{\hbar\Omega_{0}}{2}\sigma_{x}+J_{0}(\frac{\Omega_{R}}{\omega})\left(\frac{\hbar\delta_{0}}{2}+\alpha_{0}k_{x}\right)\sigma_{z},
\end{equation}
has the same functional form as the undriven Hamiltonian $H$. The
parameters $\delta_{0}$ and $\alpha_{0}$, present in $H$, have
been replaced here by their effective counterparts $\delta=J_{0}(\frac{\Omega_{R}}{\omega})\delta_{0}$
and $\alpha=J_{0}(\frac{\Omega_{R}}{\omega})\alpha_{0}$. The offset
$\Omega_{0}$ is the same in both Hamiltonians. The energy bands corresponding
to $H_{0}$ are given by

\begin{equation}
E_{\pm}(k_{x})=\frac{\hbar^{2}k_{x}^{2}}{2m}+E_{L}\pm\sqrt{\left(\frac{\hbar\Omega_{0}}{2}\right)^{2}+J_{0}^{2}(\frac{\Omega_{R}}{\omega})\left(\frac{\hbar\delta_{0}}{2}+\alpha_{0}k_{x}\right)^{2}}.\label{eq:8}
\end{equation}
$H_{0}$ provides the framework used in the design of the technique
of control: the modification in real time of the effective SOC strength
($\alpha$) and detuning ($\delta$) can be achieved by adiabatically
varying the modulation amplitude $\Omega_{R}$. 

The validity of this scheme was tested in Ref. \cite{key-SpielmanControl}.
There, the dynamics of a condensate, prepared in the lowest band,
was monitored at different values of $\Omega_{R}$. In each case,
the measurements were carried out after implementing a linear ramp
to adiabatically vary $\Omega_{R}$ from zero to its final value.
We stress that the efficiency of the control method depends strongly
on the character of the system evolution. An adiabatic following of
the lowest eigenstate of the changing Hamiltonian is needed for the
method to be efficient.

Different aspects of the comparison between the theoretical predictions
and the experimental results must be stressed. First, the values of
SOC strength, determined by directly measuring the momentum distribution,
were shown to agree with those predicted by the model. Second, the
system was also found to present heating, parametrized in terms of
the temperature in the presentation of the results. This effect was
particularly relevant in a central range of driving frequencies frequencies,
namely, for $2\,\textrm{kHz}<\omega/2\pi<7\,\textrm{kHz}$. No explanation
of these heating effects was given by the theory. However, the authors
presented a qualitative analysis, using a Floquet-state formalism,
which provided some clues on the found stability of the system at
large and small frequencies, and on the instability at intermediate
frequencies. Apart from the energy bands as function of the driving
frequency, the spatial density distributions were obtained. See the
Supplemental Material in Ref. {[}7{]}. The explanation of the heating
effects, and, in general, the departure of the system dynamics from
the picture given by $H_{0}$, is tackled in the next section.

\section{Single-particle description of the heating mechanisms}

Here, we will concentrate on general features of the heating that
can be traced to the single-particle dynamics. Although, we will basically
deal with the uniform scenario described in the previous section,
we will incidentally refer to differential aspects of the role of
the harmonic confinement. Many-body interaction effects will be incorporated
into the description further on.

\subsection{Inter-band transfer of population induced by the driving}

Crucial to the description of the heating detected in the experiments
is the inclusion in our approach of the first-order correction to
$H_{0}$, given by

\begin{equation}
H_{1}=2J_{1}(\frac{\Omega_{R}}{\omega})\sin(\omega t)\left(\frac{\hbar\delta_{0}}{2}+\alpha_{0}k_{x}\right)\sigma_{y}.
\end{equation}
It is apparent that, provided that the driving frequency is quasi-resonant
with the inter-band splitting, i.e., for $\hbar\omega\sim E_{+}(k_{x})-E_{-}(k_{x})$,
$H_{1}$ can represent an efficient mechanism of population transfer
from the lowest energy-band $E_{-}(k_{x})$ to the excited band $E_{+}(k_{x})$.
It is important to take into account that, since $H_{1}$ commutes
with $p_{x}$, the transition conserves the momentum: it corresponds
to a vertical line in a diagram of energy bands versus the quasi-momentum
$k_{x}$. When the harmonic confinement is considered, the coupling
induced by $H_{1}$ becomes more complex since the eigenstates do
not have a well-defined momentum. Given the weakly-harmonic confinement
realized in practice, a quasi-continuum approximation is feasible,
and the use of the term ``band'' is still appropriate. The following
discussion applies also to that case. 

Outside the resonance region, no significant transfer of population
between the bands is induced by $H_{1}$. In particular, for high
driving-frequencies, $\hbar\omega\gg E_{+}(k_{x})-E_{-}(k_{x})$,
the stability of the system in the lowest band is guaranteed, and,
consequently, the proposed strategy of control is sound. Indeed, $H_{0}$
provides then a satisfactory (coarse-grained) description of the dynamics:
the higher-order (oscillating) terms in the expansion of $H^{\prime}$,
which are given by 

\begin{equation}
H_{n}=2J_{n}(\frac{\Omega_{R}}{\omega})\cos(n\omega t)\left(\frac{\hbar\delta_{0}}{2}+\alpha_{0}k_{x}\right)\sigma_{i}^{(n)},\;(n\geq1)
\end{equation}
where $\sigma_{i}^{(n)}=\sigma_{z}$ when $n$ is even, and $\sigma_{i}^{(n)}=\sigma_{y}$
when $n$ is odd, can be averaged out to zero since the driving period
$2\pi/\omega$ is much smaller than any other relevant time scale
in the dynamics. The applicability of the averaging method is supported
by the fact that, as $\Omega_{R}/\omega$ decreases, the functions
$J_{n}(\Omega_{R}/\omega)$ ($n\geq1$) diminish and eventually go
to zero. In turn, as the magnitude of the perturbative terms decreases,
the associated time scale grows, allowing the coarse graining over
the driving period. In the opposite regime, namely, for $\hbar\omega\ll E_{+}(k_{x})-E_{-}(k_{x})$,
an adiabatic approximation can be implemented in the original driven
Hamiltonian, i.e., in $H$ with $\Omega_{0}$ replaced by $\Omega(t)$.
The system, initially prepared in the lowest band, evolves following
adiabatically the lowest eigenstate of $H(t)$. Consequently, no inter-band
transfer of population takes place either in this range. Note the
differences existent between this adiabatic regime, associated with
a low driving-frequency $\omega$, and that applied in the control
method, characterized by a slow variation of the driving amplitude
$\Omega_{R}$, compatible with a high driving-frequency.

From the above description, the observed emergence of instability
in the quasi-resonant region is explained as resulting from the transfer
of population to the quasi-continuum of states of the upper band.
The additional role of the decay of the excited state will be taken
into account later on. In Ref. \cite{key-SpielmanControl}, a numerical
study based on the Floquet formalism with a reduced number of states
simulated the unstable behavior as a crossing between bands. Our characterization
of the term that induces the loss of population allows identifying
precisely the spectral region where that behavior becomes more relevant.
Additionally, our analytical approach paves the way for designing
a strategy for stabilizing the system: one can think of fitting $\Omega_{R}/\omega$
to one of the zeros of $J_{1}(\Omega_{R}/\omega)$ in order to suppress
the perturbative inter-band coupling. To validate this proposal, we
must guarantee that the rest of the terms in the expansion of the
Hamiltonian do not significantly contribute to heating. This is the
case in the addressed unstable region: it is apparent that, for $\hbar\omega\sim E_{+}(k_{x})-E_{-}(k_{x})$,
those additional terms are out of resonance with the inter-band transition,
and, therefore, given their (reduced) magnitude, are irrelevant to
population transfer. In contrast, one must contemplate the possibility
of having significant heating in a series of spectral windows defined
by the conditions $\hbar n\omega\sim E_{+}(k_{x})-E_{-}(k_{x})$,
($n>1$), where the different higher-order terms $H_{n}$ are quasi-resonant
with the inter-band transition. A detailed analysis of the system
response in each case is needed. Further on, we will present results
for the magnitude of the different corrections. Still, we can anticipate
here a qualitative evaluation of the relative importance of the different
terms  by addressing a particular situation. Let us consider working
conditions where the effect of each of the terms can be expected to
be maximum; namely, we assume that, apart from fixing $\omega$ inside
the $n$-quasi-resonant region, the parameter $\Omega_{R}$ is adjusted
in order to have a maximum of $J_{n}(\Omega_{R}/\omega)$. From the
decreasing magnitude of the maxima as $n$ increases, it is inferred
that the relative importance of the different terms in the expansion
of the Hamiltonian (at their maximum efficiencies in population transfer),
diminishes as the order grows. Here, a general remark on the limitations
of the strategy proposed for stabilizing the system is pertinent:
although the loss of population can be curbed by reducing $J_{1}^{2}(\Omega_{R}/\omega)$
via an appropriate choice of the driving parameters, this implies
also fixing the SOC parameters in the zero-order Hamiltonian, i.e.,
loosing the possibility of steering the dynamics.

\subsection{Heating induced by departures from adiabaticity}

In the above analysis, a fixed $\Omega_{R}$, i.e., a concluded adiabatic
ramp of $\Omega_{R}$, was assumed. We turn now to deal with a more
realistic scenario by explicitly incorporating the ramping process
into our approach. In the experiments, a linear ramp was realized;
accordingly, here, we consider $\Omega_{R}(t)=\Omega_{R}t/T$, where
$T$ denotes the length of the ramp. Moreover, we work in the representation
of states with fixed \emph{pseudospin} component and time-dependent
external part, specifically, a plane wave with quasi-momentum following
the changing band-minimum. Again, we stress that, although the characterization
of the eigenstates becomes more elaborate when harmonic confinement
is considered, the following qualitative arguments are equally valid.
The corresponding corrections to the Hamiltonian are obtained by including
the specific time dependence of the modulation amplitude in the unitary
transformation, i.e., by replacing $\Omega_{R}$ by $\Omega_{R}(t)$
in Eq. (\ref{eq:3}), and, additionally, by taking into account the
variation of the applied basis. The transformed Hamiltonian, obtained
from Eq. (4), is given by a more involved expression, which includes
the non-adiabatic terms 
\begin{equation}
H_{na}=-\frac{\hbar\Omega_{R}}{2T\omega}\sin(\omega t)\sigma_{x}-\frac{\hbar k_{L}\Omega_{R}}{2T\omega}J_{1}(\Omega_{R}t/\omega T)x\sigma_{z},\: t\leq T,\label{eq:11}
\end{equation}
where $x$ is the operator conjugate to $p_{x}$, and, for simplicity,
we have taken the particular set of SOC parameters $\delta_{0}=0$
and $\Omega_{0}=0$. The first term comes from the explicit time dependence
of $U(t)$ introduced by $\Omega_{R}(t)$. It has a structure similar
to that of the previously analyzed driving term, i.e., of that resulting
from incorporating the modulation of the Raman amplitude given by
Eq. (2) into Eq. (1). Indeed, in the quasi-resonant regime {[}$\hbar\omega\sim E_{+}(k_{x})-E_{-}(k_{x})${]},
this term (like $H_{1}$) can lead to heating via interband transfer
of population. However, for the considered parameters and ramping
rate, the magnitude of the associated effect is $\sim20$ times smaller
than that of the original driving. The second term in Eq. (11) is
rooted in the time-dependent character of the motional component of
the adiabatic states. Specially relevant is its dependence on $x$,
which corresponds to a \emph{pseudospin}-differentiated\emph{ dipolar}
structure. Because of the form of its time dependence, entirely contained
in the argument of the Bessel function, this term cannot resonantly
couple the bands. However, it can lead to heating via the generation
of intra-band excitations. In this sense, it can account for the low-frequency
oscillations, observed in the experiments, which were clearly differentiated
from the main heating process. In Ref. \cite{key-SpielmanControl},
the origin of those features was already traced to departures from
the intended adiabatic following of the ground state induced by imperfections
in the realization of the slow ramp. Those oscillations are properly
described as collective modes in a many-body approach; their persistence
 at relatively large times will be analyzed further on. Note that
both terms in Eq. (\ref{eq:11}) consistently vanish in the limit
of large ramp-length $T$.

\subsection{A proposal for evaluating the heating rate applying the Fermi's Golden
Rule}

As a first step in the simulation of the experimental results for
the heating rate, we propose the analysis of the population transfer
from the initial lowest-band state $\left|\varphi^{(i)}\right\rangle $
to the final upper-band state $\left|\varphi^{(f)}\right\rangle $
in terms of the Fermi's Golden Rule. Accordingly, after expressing
the time-dependent perturbation as  $H_{1}=W\sin(\omega t)$, where
$W\equiv2J_{1}(\frac{\Omega_{R}}{\omega})\left(\frac{\hbar\delta_{0}}{2}+\alpha_{0}k_{x}\right)\sigma_{y}$
is a time-independent operator {[}we consider that the ramp is slow
enough to neglect the contribution to interband transfer coming from
the first term in Eq. (11){]}, the transition rate $\Gamma(\omega)$
is written as \cite{key-CohenBookQuantumM}

\begin{eqnarray}
\Gamma(\omega) & = & \frac{\pi}{2\hbar}\sum_{f}\left|\left\langle \varphi^{(f)}\left|W\right|\varphi^{(i)}\right\rangle \right|^{2}\delta(E^{(f)}-E^{(i)}-\hbar\omega),\label{eq:12}
\end{eqnarray}
 with $E^{(i)}$ and $E^{(f)}$ standing, respectively, for the energies
of the initial and final states. To evaluate $\Gamma$, we must first
identify the states $\left|\varphi^{(f)}\right\rangle $ which, in
addition to being connected with $\left|\varphi^{(i)}\right\rangle $
by the perturbation $W$, have energies fulfilling the restriction
$E^{(f)}=E^{(i)}+\hbar\omega$. Given the weakly-harmonic confinement
realized in practice, a quasi-continuum approximation can be applied
to sum the contributions of the set of possible final states. Then,
from the form of the density of states and from the dependence of
the matrix element $\left\langle \varphi^{(f)}\left|W\right|\varphi^{(i)}\right\rangle $
on the different state parameters, the heating rate can be evaluated.
Here, the importance of incorporating the harmonic trap into the model
must be emphasized. The differences with the uniform scenario are
particularly relevant to the evaluation of the required matrix elements.
In the uniform case, the transition takes place only at exact resonance
between the external frequency and the interband separation. Consequently,
$\Gamma(\omega)$ is predicted to be infinitely sharp at this resonance
frequency. When the harmonic confinement is considered, the momentum
distribution of the ground state widens, and, then, the occurrence
of transitions at frequencies different from the resonant one is predicted,
(see Fig. 1). Still, a more detailed analysis of the functional form
of $\Gamma(\omega)$ is required to explain the results of the experiments,
carried out in diverse frequency regimes. For such purpose, additional
fundamental aspects of the system, like atom-interaction characteristics,
(variations in the form of the ground state, and, damping effects
associated with the nonstationary character of the excited states),
must be taken into account. Moreover, it is also necessary to tackle
the modification of the system parameters that takes place in the
ramp implemented to adiabatically vary $\Omega_{R}$ from zero to
its final value. The evaluation of the losses in the ramp, where the
rate is actually a function of time, is necessary. To incorporate
those elements into our framework, we must generalize the above formulation
of the Fermi's Golden Rule. This is the objective of the next two
sections.

\section{Atomic-interaction effects in the heating processes}

\subsection{Effects associated with ground-state characteristics}

The description of the atomic-interaction effects is required by the
experimental conditions, which correspond to the preparation as a
condensate. Interaction effects primarily appear in the initial-state
wave-function, which becomes dependent on the interaction strength
and on the number of particles, and in the excited-mode characteristics,
in particular, in the functional form of the dispersion relations.
Due to the SOC, the system presents specific components. In particular,
there are different interaction strengths associated with the spin
combinations. They will be denoted as $g_{\eta,\eta^{\prime}}$, where
$\eta$ and $\eta^{\prime}$ refer to the spins. As shown in previous
studies, depending on the SOC characteristics and on the values of
$g_{\eta,\eta^{\prime}}$, the form of the wave-function of the ground
state can display a varied topology \cite{key-StringariSOCreview}.
In the case of uniform confinement, three phases, known as stripe,
plane-wave (separate dressed-state), and single-minimum phases, can
appear. Actually, the description of the system is quite complex in
a general regime. Here, we will concentrate on a particular set of
SOC parameters, which will allow us to present a simplified picture
of the role of interaction effects in heating. Specifically, we will
fix the detuning and the offset of the Raman-laser arrangement to
the values $\delta_{0}=0$ and $\Omega_{0}=0$. These parameters along
with the scattering lengths for the considered system correspond to
the regime identified as separate (single-dressed state) phase, which
defines the condensation in a single plane-wave state, the sign of
the quasimomentum being determined by the particular set of $g_{\eta,\eta^{\prime}}$
\cite{key-StringariSOC1,key-ZhangPhase}. This restriction, which
will be considered also in the analysis of the elementary excitations,
is conformable with our aim of identifying mechanisms of general relevance. 

The description of the coupling between states, and, in turn, of the
whole transfer process can be significantly altered when interaction
characteristics are taken into account. It is worth pointing out that
the factorization of the initial wave-function and the subsequent
reduction in dimensionality applicable to a single-particle system
are precluded in the many-body case. Given the form of the coupling
term $W$, the distribution of momenta of the initial and final states
plays an essential role in the population transfer. Moreover, since
some of the experiments were carried out in the high-frequency regime,
the characterization of the high-momentum region is particularly important.
Note that the analysis of this region is not frequent in studies of
condensates. These arguments are illustrated in Fig. 1, where results
for the heating rate as a function of the driving frequency are presented.
For simplicity, an isotropic harmonic trap has been considered. In
particular, we have taken the three trap frequencies $\omega_{x}$,
$\omega_{y}$, and $\omega_{z}$, equal to the geometric mean of the
frequencies corresponding to the experimental setup \cite{key-SpielmanControl}.
The predictions of the Thomas-Fermi model substantially differ from
those of a single-particle approach. Whereas, in the single-particle
case, the form of $\Gamma(\omega)$ points to a strong localization
of the population transfer in the resonance region, the Thomas-Fermi
model accounts for significant heating even at large frequencies.
Given the variety of ground-state forms that can emerge as the SOC
parameters are changed, it is pertinent to check the robustness of
the above findings when the modeling of the ground state is modified.
Indeed, as anticipated, we have found that the form of $\Gamma(\omega)$
is strongly dependent on the ground-state characteristics that affect
the high momenta. Even so, at this point, we cannot conclusively trace
the form of the experimental heating curve to that component of the
system: in the following, we will see that the observed widening of
$\Gamma(\omega)$ can be rooted also in damping of the excited state.
In fact, it will be uncovered that strong damping can make irrelevant
the precise modeling of the ground state. To evaluate the relative
magnitude of both effects, specifically targeted experiments are required.
Note that, in Fig. 1, as in other figures in the article, we have
omitted the range of low frequencies where the applied perturbative
scheme fails, i.e., when the initially applied unitary transformation
and the subsequent averaging procedure do not lead to a useful reduction
of the system. 

\begin{figure}[H]
\centerline{\includegraphics{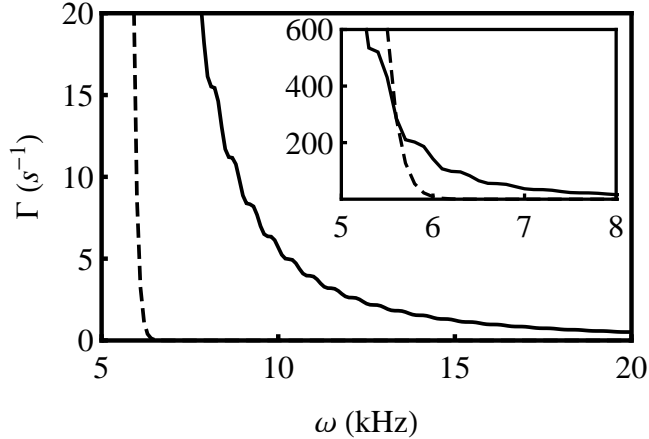}}\caption{Heating rate as a function of the driving frequency for $\Omega_{R}/\omega=1$.
The ground state corresponds to the Thomas Fermi approximation (continuous
line), and, (in a single-particle description), to the ground state
of a harmonic oscillator (dashed line). In the inset the vertical
scale is enlarged.}
\end{figure}

\subsection{The decay of the excited states: characterization of the elementary
excitations}

The nonstationary character of the excited state must be taken into
account. In the description that includes atomic interaction effects,
the final state in the interband transition will be shown to be embedded
in the set of elementary excitations of the condensate, and, coupled
to them. As a result, a damping mechanism, standardly known as \emph{Beliaev}
or \emph{Landau} \emph{damping}, turns up; it basically consists in
nonlinear mode-coupling, i.e., in interaction between modes rooted
in scattering terms of order higher than quadratic, (not included
in the mode characterization). We proceed to trace this mechanism
in our system.

We will depart from the diagonalization of a quadratic approximation
to our many-particle zero-order Hamiltonian, i.e., to the generalization
of $H_{0}$ obtained by adding the inter-atomic potential \cite{key-StringariBook,key-PethickBook}.
Subsequently, we will build a perturbative scheme, where corrections
to the quadratic setup can be incorporated. Those additional terms
account for coupling between the unperturbed states. The resulting
\emph{mixing} of modes will be introduced into an operative procedure
to evaluate the heating rate through the modification, by a shift
and a width, of the energy of the state characterized as $\left|\varphi^{(f)}\right\rangle $
in Eq. (\ref{eq:12}). The shift implies the redefinition of the resonance,
and the width incorporates the decay. In this form, the description
given by Eq. (\ref{eq:12}) is extended in the  unperturbed-state
representation. Alternatively, one could switch to the picture of
eigenstates of the perturbed system and regard the \emph{mixing} as
leading to the enlargement of the range of the states that can be
reached from the initial state through $H_{1}$. Then, damping should
be incorporated into Eq. (\ref{eq:12}) via the modification of the
coupling matrix-element and of the effective density of states \cite{key-CohenBookAtomPhoton}.

Useful to isolate the basic components of damping is to tackle a simplified
model. We will consider a uniform system. The effects of the harmonic
confinement will be discussed later on. Additionally, the interaction
potential will be simply characterized by the scattering lengths,
as in usual ultra-cold-atom contexts. The driving-dependent scattering
terms that emerge from the application of the unitary transformation
given by Eq. (3) will be analyzed further on. By now, we deal only
with the (dominant) \emph{static} scattering potential. For the considered
set of SOC parameters, ($\delta_{0}=0$ and $\Omega_{0}=0$), the
expression for the energy bands, obtained from Eq. (\ref{eq:8}),
reads

\begin{equation}
E_{\pm}(k_{x})=\frac{\hbar^{2}(k_{x}\mp k_{D})^{2}}{2m}+E_{L}+E_{D},
\end{equation}
 where we have defined $k_{D}=\frac{\alpha m}{\hbar^{2}}$ and $E_{D}=-\frac{\hbar^{2}k_{D}^{2}}{2m}$.
We recall that $\alpha=J_{0}(\frac{\Omega_{R}}{\omega})\alpha_{0}=J_{0}(\frac{\Omega_{R}}{\omega})\frac{2E_{L}}{k_{L}}$.
Then, the complete system can be regarded as composed by two uniform
subsystems displaced in different directions in the quasi-momentum
quantity $k_{D}$ (see Fig. 2).

\begin{figure}[H]
\centerline{\includegraphics{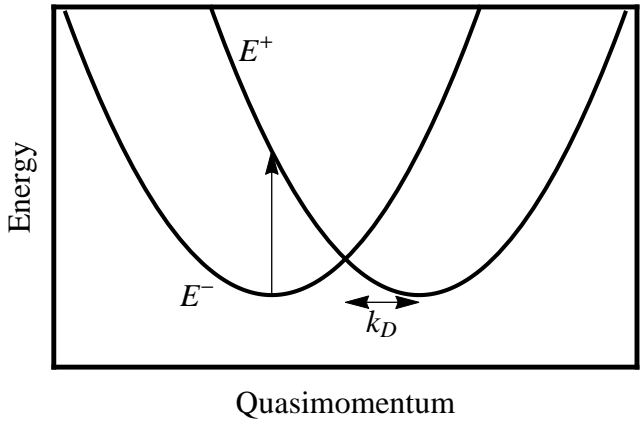}}\caption{Diagram of energy bands versus quasi-momentum. ($\delta_{0}=0$ and
$\Omega_{0}=0$).}
\end{figure}

As corresponds to the assumed SOC parameters and to the values of
$g_{\eta,\eta^{\prime}}$ for the referred realization, we consider
a condensate in the (left) band $E_{-}(k_{x})$. The term $H_{1}$
induces transitions to states in the other (right) band $E_{+}(k_{x})$,
which, initially, is not populated. Those transitions are illustrated
in Fig. 2. The final states are denoted as $\left|\varphi_{+,\mathbf{\boldsymbol{\mathbf{k}}}}^{(f)}\right\rangle $,
i.e., we specify the well-defined momentum (a uniform confinement
is being considered), and the band label. Because of the preparation
and the simplified band-structure, the elementary excitations can
be described via a direct generalization of those corresponding to
the  non-SOC scenario. The Hamiltonian is written in terms of the
operators $a_{\eta,\mathbf{k}}^{\dagger}$ ($a_{\eta,\mathbf{k}}$),
which correspond to the creation (annihilation) of a particle in the
state $\left|\eta,\mathbf{k}\right\rangle $, where $\eta$ stands
for the band (+, -) {[}(right, left){]} and $\mathbf{k}$ denotes
the quasi-momentum. The characterization of the eigenstates of the
right subsystem is complete at this point: since a single-particle
description is feasible, the operator $a_{+,\mathbf{k}}^{\dagger}$
simply corresponds to the creation of a particle in the state $\left|\varphi_{+,\mathbf{k}}\right\rangle $.
In contrast, for the left-band subsystem, our procedure continues:
the Bogolubov approach is applied to obtain the quasiparticles \cite{key-StringariBook,key-PethickBook}.
This separate characterization is possible since there are no (interband)
terms in the quadratic approximation. We will see that it is from
the cubic corrections that interband coupling emerges. Displaced quasi-momenta
in the $x$-direction are used, i.e., we take in the left band ($-$)
$\tilde{k}_{x}=k_{x}+k_{D}$, $\tilde{k}_{y}=k_{y}$, $\tilde{k}_{z}=k_{z}$,
and write

\begin{eqnarray}
b_{-,\tilde{\mathbf{k}}} & = & u_{\tilde{k}}a_{-,\tilde{\mathbf{k}}}+v_{\tilde{k}}a_{-,-\tilde{\mathbf{k}}}^{\dagger}\\
b_{-,-\tilde{\mathbf{k}}}^{\dagger} & = & u_{\tilde{k}}a_{-,-\tilde{\mathbf{k}}}^{\dagger}+v_{\tilde{k}}a_{-,\tilde{\mathbf{k}}}
\end{eqnarray}
 where $b_{-,\tilde{\mathbf{k}}}^{\dagger}$($b_{-,\tilde{\mathbf{k}}}$)
denotes the creation (annihilation) operator of a quasi-particle,
and $u_{\tilde{k}}$ and $v_{\tilde{k}}$ are real coefficients, which
are standardly obtained. The associated (unperturbed) energies are
given by

\begin{equation}
E_{-,\tilde{k}}=\sqrt{\frac{(\hbar\tilde{k})^{2}}{2m}\left[\frac{(\hbar\tilde{k})^{2}}{2m}+2\rho g\right]}
\end{equation}
where $\rho=N/\mathcal{V}$ is the particle density, ($N$ denotes
the number of particles in the condensate and $\mathcal{V}$ is the
effective volume of confinement). Moreover, $g$ denotes the interaction
strength of the atoms in the condensate. It is $g_{\downarrow\downarrow}\equiv g$
that appears in Eq. (16), as we are describing the left band, where
the condensate has been prepared. 

The above reduction of the system is feasible because of the simplifications
applicable to the considered set of SOC parameters. For a generic
regime, the characterization of the elementary excitations is a much
more involved problem.

\subsection{The effect of nonlinear mode-coupling}

The eigenstates of the quadratic Hamiltonian for the left-band subsystem,
characterized by the operators $b_{-,\tilde{\mathbf{k}}}$, are actually
coupled through the cubic and quartic terms, left out in the above
(zero-order) approximation. They are also coupled to the eigenstates
of the right-band subsystem, i.e., to the single-particle states $\left|\varphi_{+,\tilde{\mathbf{k}}}\right\rangle $.
We have changed their labels by using the displaced momenta. The same
notation will be applied to the corresponding operators $a_{+,\tilde{\mathbf{k}}}^{\dagger}$
($a_{+,\tilde{\mathbf{k}}}$). Since we are interested in the decay
of the states $\left|\varphi_{+,\tilde{\mathbf{k}}}\right\rangle $,
we will focus on the (leading) cubic terms that include them. In second
quantization, those terms have the generic  form

\begin{equation}
V_{int}^{(3)}\sim\sqrt{N}g_{\uparrow\downarrow}(u_{\tilde{k^{\prime}}}-v_{\tilde{k^{\prime}}})b_{-,\tilde{\mathbf{k}}^{\prime}}^{\dagger}a_{+,\tilde{\mathbf{k}}-\tilde{\mathbf{k}}^{\prime}}^{\dagger}a_{+,\tilde{\mathbf{k}}},
\end{equation}
which defines a two-particle scattering where one of the input particles
is in $\left|\varphi_{+,\tilde{\mathbf{k}}}\right\rangle $ and the
other in the condensate, the output states being $\left|\varphi_{+,\tilde{\mathbf{k}}-\tilde{\mathbf{k}}^{\prime}}\right\rangle $
and the quasi-particle created by $b_{-,\tilde{\mathbf{k}}^{\prime}}^{\dagger}$.
($g_{\uparrow\downarrow}$ is the interaction strength relevant to
that process). Momentum conservation has been incorporated into the
notation; energy conservation implies an additional restriction on
the particle momenta. The factor $\sqrt{N}$ appears in Eq. (17) because
one of the interacting particles is in the condensate. Since the quartic
terms correspond to scattering with no particles in the condensate,
the magnitude of their correction to the heating rate can be anticipated
to be a factor $N$ smaller than that associated with $V_{int}^{(3)}$. 

The extension of the quadratic setup by incorporating cubic terms
leads to the decay of the unperturbed states \cite{key-Beliaev1,key-Beliaev2,key-RonenBeliaev,key-Sirlin}.
\emph{Beliaev} and \emph{Landau} \emph{damping} processes are variations
of this physical mechanism \cite{key-StringariDamping,key-FedichevDamping,key-RonenBeliaev,key-giorgini}.
In \emph{Beliaev damping}, an excited mode decays into two lower ones.
In \emph{Landau damping}, an upper state is reached via collisions
between thermally-populated lower modes. The relative importance of
those processes depends on the temperature and the density of accessible
output modes. In the scenario considered here, the highly-excited
character of the decaying state implies the existence of a large number
of accessible output modes, and, consequently, the potentially important
role of \emph{Beliaev damping}. Moreover, because of the very low
temperature considered, the initial thermal populations of excited
modes can be neglected, and, therefore, the secondary character of
\emph{Landau damping} can be conjectured.

Through a standard perturbative treatment, the effect of the terms
$V_{int}^{(3)}$, (in a Beliaev-damping scenario), is incorporated
into our description as a modification of the energy of the final
state, which is is written as 
\begin{equation}
\mathcal{E}_{-}(k_{x})=\frac{\hbar^{2}k_{x}^{2}}{2m}+E_{L}+E_{D}+\delta_{k_{x}}-i\hbar\frac{\gamma_{k_{x}}}{2}.
\end{equation}
Here, the unperturbed value appears corrected by the effective width
$\gamma_{k_{x}}$ and shift $\delta_{k_{x}}$, which can be evaluated
analytically in a highly-diluted regime, i.e., for $(\rho a^{3})^{1/2}\ll1$,
where $a$ corresponds to any of the involved scattering lengths. 

The modification of Eq. (12) to incorporate the decay of the final
state is straightforward \cite{key-CohenBookAtomPhoton}. In the case
of a uniform model for the ground state, the heating rate is given
by 

\begin{eqnarray}
\Gamma(\omega) & = & \pi J_{1}^{2}(\frac{\Omega_{R}}{\omega})(\alpha_{0}k_{D})^{2}\frac{\gamma_{k_{x}}}{\left[\Delta E(k_{D})-\hbar\omega\right]^{2}+\left(\hbar\frac{\gamma_{k_{x}}}{2}\right)^{2}}.
\end{eqnarray}
{[}$\Delta E(k_{D})\equiv E_{+}(-k_{D})-E_{-}(-k_{D})${]}. The effect
of the shift $\delta_{k_{x}}$has been neglected: due to the high
frequency of the driving, the shift in the energy hardly modifies
the decay rate. The generalization of the above expression to a nonuniform
model for the ground state is direct. Damping obtained using appropriate
values for the density and scattering lengths in the system is incorporated
into Figs. 3 to 8. 

Our results for damping at high energies, where the elementary excitations
have single-particle character, can be reasonably expected to be robust
when trapping effects are incorporated. In that range, the uniform
approximation can be assumed to incorporate the essential components,
the harmonic effects being expected to merely modify the form of the
states, and, consequently, the coupling matrix elements. In contrast,
the trapping conditions cannot be avoided in the analysis of the collective
modes generated by the imperfect adiabatic character of the ramp.
The distribution of frequencies is determined by the trap characteristics.
The observed persistence of those low-frequency oscillations can be
understood taking into account their small damping rate. Indeed, in
the low energy-regime, where the elementary excitations have collective
character, \emph{Beliaev damping} is negligible: the number of accessible
(output) modes is significantly reduced due to the restrictions that
the discrete character of the spectrum adds to the conservation of
energy and momentum. It is \emph{Landau damping}, proportional to
the temperature in the considered low-temperature regime, that can
be assumed to account for the observed (weak) damping.

The presence of $\gamma_{k_{x}}$ in Eq. (19) indicates that the decay
can significantly affect the width of the instability region. This
is reflected in Fig. 3, where $\Gamma(\omega)$, obtained using different
models, is represented. Actually, the appearance of the damping effects
varies with the ground-state characteristics: it is the relative importance
of damping and widening due to the momentum distribution that determines
the form of the curves. Namely, the predictions, presented in Fig.
1, for the (nondissipative) Thomas-Fermi model are slightly modified
by damping. In contrast, significant changes take place when variations
of the ground-state model are considered: damping appreciably broadens
the rate corresponding to characterizations of the ground state as
a plane wave and as a harmonic-oscillator state. Moreover, in those
cases, where the role of damping in widening the function $\Gamma(\omega)$
dominates over the effect of the momentum distribution, the precise
modeling of the ground state becomes irrelevant, as can be seen in
Fig. 4. Special attention requires the analysis of the results for
$\Gamma(\Omega_{R}/\omega)$. The central importance of the reduced
amplitude $\Omega_{R}/\omega$ in our description is reflected in
the curves in Figs. 4 and 5. The separate presentation of the results
for the plane wave and harmonic-oscillator ground state (Fig. 4) and
those of the Thomas-Fermi distribution (Fig. 5) is imposed by the
widely different magnitude of them. In passing, those results emphasize
the need of including atomic-interaction effects to account for the
experimental findings. We must clearly separate in those curves the
features that are specific to the used set of parameters from those
of general relevance. Particularly important is to trace the appearance
of zeros in the heating rate. As formerly discussed, in a general
regime, it is the factor  $J_{1}^{2}(\Omega_{R}/\omega)$ that is
certainly known to determine the magnitude of heating. Then, irrespective
of the set of parameters, the oscillations of $J_{1}(\Omega_{R}/\omega)$
can be expected to be relevant to the form of $\Gamma(\Omega_{R}/\omega)$.
In particular, the zeros of $J_{1}(\Omega_{R}/\omega)$ trivially
lead to the inhibition of the population transfer in a first-order
perturbative treatment. The global form of the curves of $\Gamma(\Omega_{R}/\omega)$
presented in Fig. 4 indicates the emergence of additional effects,
rooted in the form of the coupling matrix-elements. The additional
zeros that appear in $\Gamma(\Omega_{R}/\omega)$ are specific to
the particular set of parameters considered: they correspond to zeros
of $J_{0}(\Omega_{R}/\omega)$, as can be checked by an approximate
analytic evaluation of the matrix elements. They cannot be used to
curb the effects of heating simply because, for the corresponding
driving parameters, there is no SOC in the zero-order Hamiltonian. 

\begin{figure}[H]
\centerline{\includegraphics{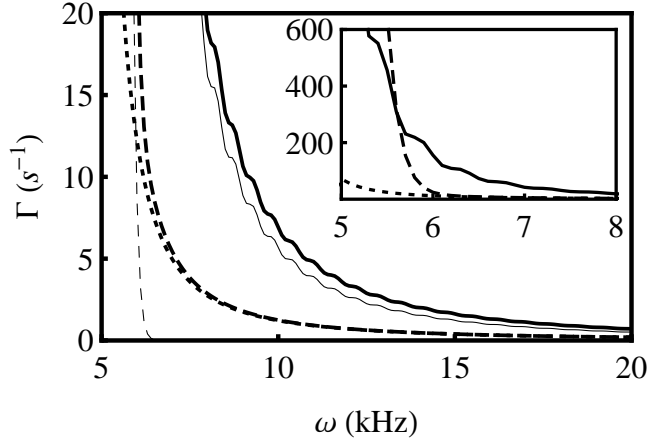}}\caption{Heating rate as a function of the driving frequency for $\Omega_{R}/\omega=1$.
The ground state corresponds to the Thomas Fermi approximation (continuous
line), to the ground state of a harmonic oscillator (dashed line),
and to a plane wave (dotted line). Thick lines incorporate damping,
thin lines correspond to the nondissipative system (data from Fig.
1). In the inset the vertical axis is enlarged.}
\end{figure}

\begin{figure}[H]
\centerline{\includegraphics{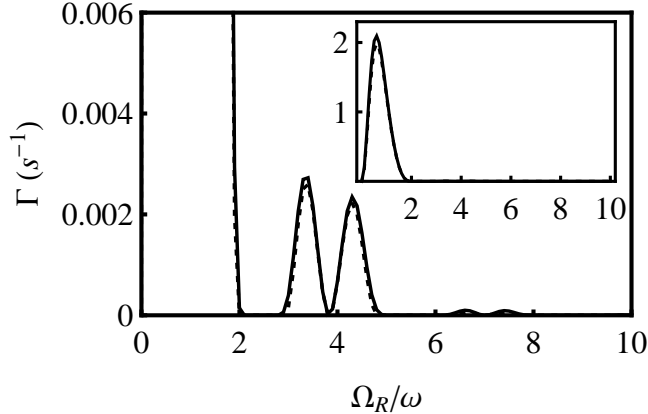}}\caption{Heating rate as a function of the reduced driving amplitude $\Omega_{R}/\omega$
for $\omega=\textrm{10 kHz}$. The ground state has been modeled as
a plane wave (continuous line) and as the ground state of a harmonic
oscillator (dashed line). Damping has been included. In the inset
the vertical axis is enlarged.}
\end{figure}

We have assumed that our first-order approach provides a sufficiently
sound framework to calculate the heating rates. This has been confirmed
by evaluating the corrections introduced by higher-order perturbative
terms, i.e., by $H_{n}$ with $n>1$. Fig. 5 provides a global view
of the magnitude of those corrections. There, the validity of our
framework is apparent. In our Figures, we have fixed the coupling
strengths and the density to appropiate values. In fact, the dependence
of the population loss and instability region on those parameters
is evident: it can be simply traced back to the form of the cubic
terms given by Eq. (17). In Fig. 6, results for $\Gamma$ as a function
of $\Omega_{R}/\omega$, at three different frequencies, are presented.
The decrease of the heating rate with increasing frequencies in the
whole range for the reduced amplitude is evident. Hence, consistently
with the experimental findings, the applicability of the scenario
of control proposed in Ref. {[}7{]} is shown to be recovered in the
high-frequency regime.

\begin{figure}[H]
\centerline{\includegraphics{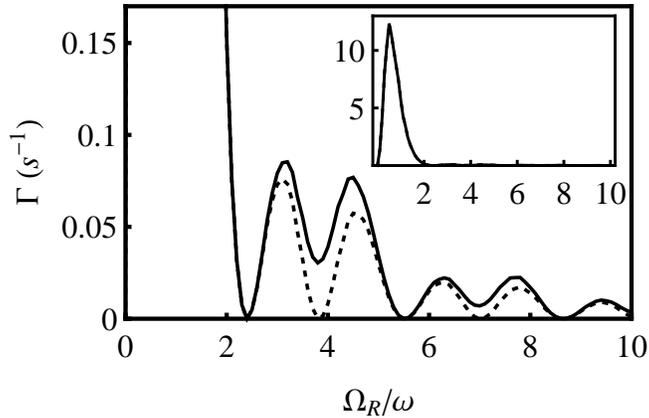}}\caption{Heating rate as a function of the reduced driving amplitude $\Omega_{R}/\omega$
for $\omega=\textrm{10 kHz}$, obtained including only the dominant
$H_{1}$ correction in the perturbative scheme (dashed line; data
from Fig. 4) and with all corrections $H_{n}$ (continuous line).
The ground state corresponds to the Thomas-Fermi approximation. Damping
has been included. In the inset the vertical axis is enlarged.}
\end{figure}

\begin{figure}[H]
\centerline{\includegraphics{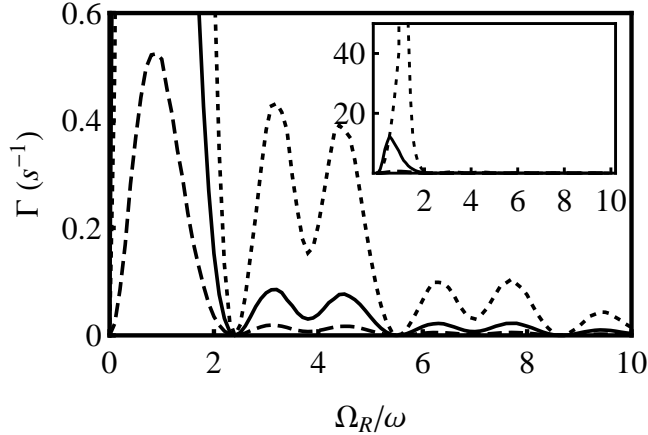}}\caption{Heating rate as a function of the reduced driving amplitude $\Omega_{R}/\omega$
at three different frequencies: $\omega_{1}=\textrm{5 kHz}$ (dotted
line), $\omega_{2}=\textrm{10 kHz}$ (continuous line), and $\omega_{3}=\textrm{20 kHz}$
(dashed line). The ground state corresponds to the Thomas-Fermi approximation.
Damping and all $H_{n}$ corrections have been included. In the inset
the vertical axis is enlarged.}
\end{figure}

\subsection{Dressing of the scattering potential}

We turn to assess the relevance of having different interaction strengths,
associated with the spin combinations. The spin-dependent scattering
potential corresponding to the experimental setup, characterized by
strengths which fulfill $g_{\uparrow\uparrow}=g_{\downarrow\uparrow}\neq g_{\downarrow\downarrow}$,
is not invariant under the unitary transformation given by Eq. (3).
In fact, the application of $U(t)$ to the interatomic potential leads
to the emergence of driving-dependent terms which must be added to
the static ones considered by now in the transformed many-body Hamiltonian.
In particular, for the formerly considered SOC parameters and system
preparation, the correction reads 
\begin{equation}
\frac{\Delta g}{4}\rho\sin[\zeta(t)](1-\cos[\zeta(t)])\sigma_{y},
\end{equation}
 where $\Delta g$=$g_{\uparrow\uparrow}-g_{\downarrow\downarrow}$.
This term accounts for interband coupling, which can be relevant,
again, in the quasiresonant case. Consequently, it can lead to heating.
Note that the structure of the correction is similar to that of $H_{1}$,
except in the absence of the dependence on $k_{x}$. Still, the induced
transitions are vertical in momentum. We have evaluated the associated
heating effects and found that, for the considered experimental conditions,
they are orders of magnitude lower than those corresponding to $H_{1}$.
We leave for future work the analysis of the potential control of
this term via the appropriate choice of system and driving parameters
and the use of Feshbach resonances. Note that the analyzed correction
is an example of how the dressing of the atomic system by driving
fields, (in the considered case, by the Raman-laser modulation), can
modify the scattering characteristics. The same physics underlies
the density-dependent synthetic gauge fields, recently found in a
related context \cite{key-LSantos}.

\section{Heating in the ramping process}

We must incorporate into our description the modification of the parameters
of $H_{0}$ that takes place in the ramp implemented to adiabatically
vary $\Omega_{R}$ from zero to its final value. First, one must deal
with  the changing inter-band energy, and, in turn, with the varying
character of the resonance between the driving frequency and the inter-band
transition. Additionally, we must take into account that, the operator
$W$, which depends on $\Omega_{R}$, varies during the ramping process.
The potential changing magnitude of damping in different segments
of the ramp must also be assessed. In the stages which correspond
to a low-energy regime, damping is irrelevant: the \emph{Beliaev processes}
are blocked because of the reduced number of accessible output states.
It is in the stages where significant inter-band splitting has been
reached, that the (Beliaev) damping becomes important. As formerly
stated,\emph{ Landau damping} is negligible due to the assumed low
temperature. From the combination of those factors, the transition
rate $\Gamma$ can present a nontrivial time-dependence, which can
allow a variety of behaviors ranging from the absence of significant
interband transfer in the whole ramp to strong heating in specific
time intervals. This is shown in Fig. 7, where we depict the heating
rate as a function of both, the frequency $\omega$, and the reduced
amplitude $\Omega_{R}/\omega$. Cases a) and b) respectively correspond
to modeling the ground state in the Thomas-Fermi approximation and
as a harmonic-oscillator state. A ramp corresponds to a fixed $\omega$
and to an increasing $\Omega_{R}/\omega$. A first noticeable feature
is, again, the robustness of the control strategy at high frequencies:
losses are negligible along the whole ramp. As the frequency diminishes,
heating becomes relevant, basically in an initial region for the (increasing)
parameter $\Omega_{R}/\omega$. After that region, the population
of the system in the lowest band remains stable. We recall that the
zeros in $\Gamma(\Omega_{R}/\omega)$ rooted in those of $J_{1}(\Omega_{R}/\omega)$
are specifically associated with the inhibition of the population
transfer. In contrast, the additional zeros that appear in $\Gamma(\Omega_{R}/\omega)$
can be traced to the considered particular set of SOC parameters;
as previously indicated, since they derive from zeros of $J_{0}(\Omega_{R}/\omega)$,
they correspond to a suppression of the SOC. Then, they cannot be
used in a strategy for curbing heating in a general regime. An additional
feature of the curves must be emphasized. Namely, at very low frequencies,
the whole perturbative scheme breaks down: the zero-order Hamiltonian
does not provide an appropriate framework for analyzing the dynamics. 

To uncover the actual magnitude of heating we must integrate the losses
along the complete ramp. In Fig. 8, we depict the corresponding energy
gained in the band transfer as a function of the driving frequency,
i.e., for different complete ramps. The ground state has been alternatively
modeled using the Thomas-Fermi approach and as the one-particle Gaussian
ground state. In the presentation of the experimental results, the
heating was parametrized in terms of an effective temperature. Here,
we do not tackle the thermalization of the system. In fact, this is
a nontrivial problem in this context. Hence, we have opted for evaluating
the heating directly from the gained energy. Our curves are shown
to peak at the resonance-frequency region. Differences in the estimated
width of the peak are rooted in the different distribution of momentum
corresponding to the applied models. The results reveal that the system
preparation can play a crucial role in heating: for ramps corresponding
to a broad range of frequencies, significant losses of population
take place before reaching the final value of $\Omega_{R}$. In the
experiments, the heating was observed to reach a plateau at large
frequencies. That feature is not reproduced in our approach, which
focuses here on heating by interband transitions. We conjecture that
feature to be rooted in additional heating mechanisms. The need of
pondering these effects in the design of the strategy of control is
apparent. An interesting subject of future work can be the design
of alternative and more effective forms of ramping, where the implications
of our analysis can be incorporated. Here, one can anticipate that
a compromise must be reached between minimizing the relevance of the
region of significant losses, for instance, by ramping faster there,
and, on the other hand, curbing the corrections associated to the
consequent departures from adiabaticity. The optimization of the ramp
is out of the scope of the present paper.

\begin{figure}[H]
\centerline{\includegraphics{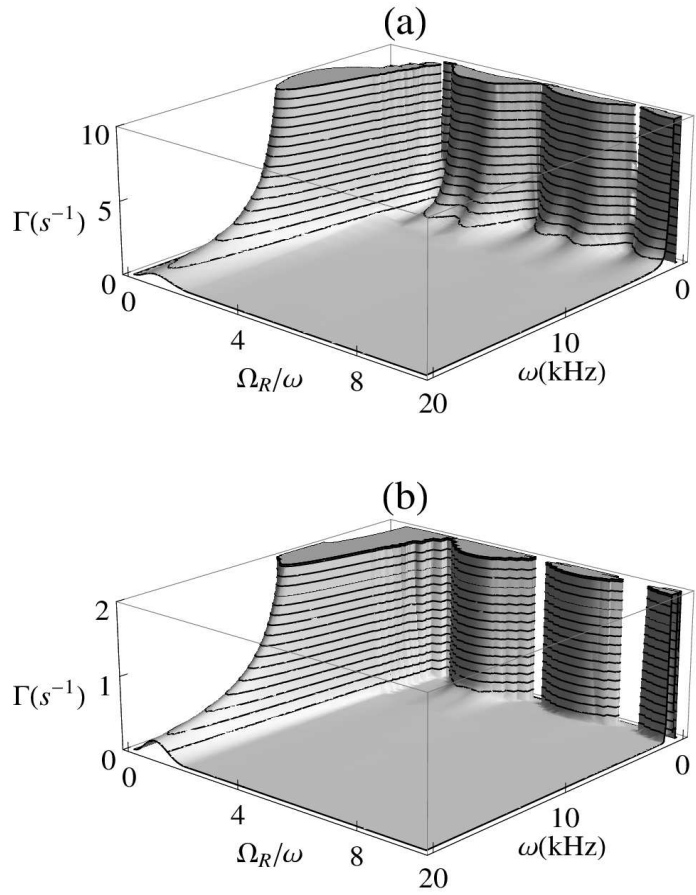}}\caption{Heating rate as a function of the driving frequency $\omega$ and
effective amplitude $\Omega_{R}/\omega$. The ground state is modeled
with the Thomas-Fermi approximation (a) and as the ground state of
a harmonic oscillator (b). Damping has been included.}
\end{figure}

\begin{figure}[H]
\centerline{\includegraphics{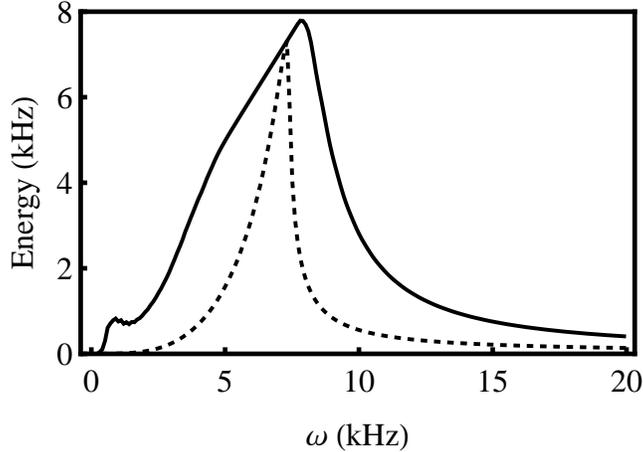}}\caption{Energy gained in the heating process as a function of the driving
frequency at $\Omega_{R}/\omega=\textrm{2.9}$. The ground state corresponds
to the Thomas Fermi approximation (continuous line) and to the ground
state of a harmonic oscillator (dashed line). Damping has been included.}
\end{figure}

\section{Concluding remarks}

The study corroborates that the restrictions of working with high
driving-frequencies and very slow variations in the modulation amplitude
are sufficient for guaranteeing the applicability of the technique
of control. In the high-frequency regime, the description of the system
given by $H_{0}$ has been found to be valid, the effect of the higher-order
corrections being then negligible. Moreover, it has been shown that
the safe range of applicability can be extended. Namely, if practical
limitations impose the use of smaller frequencies, a first general
recommendation is to avoid the region of resonance between the modulation
and the inter-band transition. This applies to any interval in the
ramping processes. Still, in the resonance region, the adjustment
of the quotient of parameters $\Omega_{R}/\omega$ to reduce the magnitude
of the Bessel function $J_{1}(\Omega_{R}/\omega)$ can be used to
curb heating. It is worth stressing, that the relevance of higher-order
corrections to the Hamiltonian of control can be enhanced by the characteristics
of the practical arrangement. For instance, in setups where a continuous
variations of the band splitting is implemented, like in realizations
of Landau-Zener analogs \cite{key-lzSOC}, different higher-order
terms can become dominant if the sequence of resonance windows is
crossed. Given the broad use of similar averaging techniques to obtain
reduced pictures in different contexts, our scheme for suppressing
the effect of the perturbations at different orders can be expected
to have quite general applicability. In the system considered, there
are no technical problems in the realization of this strategy. The
analysis has also uncovered the potential relevance of the ramp to
heating: depending on the driving frequency, significant losses of
population can occur before reaching the intended value of the driving
amplitude. In addition to curbing the inter-band transfer of population,
the methods of control must prevent the generation of intra-band excitations.
In this regard, departures from adiabaticity in the variation of the
modulation amplitude must be minimized. The development of efficient
alternative forms of the ramp which can prevent the crossing of the
resonance region and the excitation of collective modes constitutes
an interesting line for future research.

The possibility of using the driving, not only as an element of control,
but as an instrument to explore the excited modes is open. In parallel,
the development of models applicable to the trapped system can benefit
from the conclusions of our study. Future work will be dedicated to
the characterization of the elementary excitations in regimes for
the SOC parameters where the simplifications assumed in our work are
not applicable. Significant advances in this line can result from
the availability of a controllable testing ground for the theoretical
predictions. 

In the studied system, there is little difference between the scattering
strengths associated with the different spin combinations. As a consequence,
the analysis of the role of the atomic-interaction potential simplifies
considerably. However, one can think of realizations where a larger
variety of scattering lengths can be used to increase the versatility
of the scenario. In fact, our procedure has uncovered that the spin
dependence of both, the interaction potential and the driving term
in the Hamiltonian, leads to density dependent contributions to heating.
Hence, nontrivial effects can be expected from significant differences
between the involved scattering lengths. Apart from being applicable
to the Raman-laser scenario, the study presents arguments that can
be relevant  to other SOC setups \cite{key-DalibardRMP}, where dressing
of the atomic systems by classical fields is used. Moreover, since
significant characteristics of the heating have been traced to the
single-particle dynamics, one can think of extrapolating part of the
conclusions, those referring to the initial single-particle stages
of our study, to parallel arrangements in fermionic systems.

\section*{Acknowledgments}

One of us (JMGL) acknowledges the support of the Spanish Ministerio
de Economía y Competitividad and the European Regional Development
Fund (Grant No. FIS 2013-41532-P).

\end{document}